\documentclass[10pt,twoside]{article}
\usepackage{own_sngl}
\usepackage[dvips]{graphics}
\usepackage{epsfig}
\usepackage{epsf}
\usepackage[figuresleft]{rotating}
\newcommand{\oversim}[2]{\protect{\mbox{\lower0.5ex\vbox{%
   \baselineskip=0pt\lineskip=0.2ex
   \ialign{$\mathsurround=0pt #1\hfil##\hfil$\crcr#2\crcr\sim\crcr}}}}} 
\newcommand{\simgreat}{\mbox{$\,\mathrel{\mathpalette\oversim>}\,$}} 
\newcommand{\simless} {\mbox{$\,\mathrel{\mathpalette\oversim<}\,$}} 
\slugcomment{{\em {\bf ApJ, in press}}}
\lefthead{Kroupa, Burkert}
\righthead{Binary-Star Period Distribution}
\begin{document}
%
\title {On the origin of the distribution of binary-star periods}
\author {Pavel Kroupa\footnote{\small pavel@astrophysik.uni-kiel.de} 
and Andreas
Burkert\footnote{burkert@mpia-hd.mpg.de}\\
\medskip
\small{$^1$ Institut f\"ur Theoretische Physik und Astrophysik\\
Universit\"at Kiel, D-24098 Kiel, Germany\\
$^2$Max-Planck-Institut f\"ur Astronomie, K\"onigstuhl 17\\ 
D-69117 Heidelberg, Germany}}
%
\begin{abstract}
\noindent 
Pre-main sequence and main-sequence binary systems are observed to
have periods, $P$, ranging from one day to $10^{10}$~days and
eccentricities, $e$, ranging from~0 to~1. We pose the problem if
stellar-dynamical interactions in very young and compact star clusters
may broaden an initially narrow period distribution to the observed
width.  $N$-body computations of extremely compact clusters
containing~100 and~1000 stars initially in equilibrium and in cold
collapse are preformed.  In all cases the assumed initial period
distribution is uniform in the narrow range $4.5\le\,{\rm
log}_{10}P\le 5.5$ ($P$ in days) which straddles the maximum in the
observed period distribution of late-type Galactic-field dwarf
systems.  None of the models lead to the necessary broadening of the
period distribution, despite our adopted extreme conditions that
favour binary--binary interactions.  Stellar-dynamical interactions in
embedded clusters thus cannot, {\it under any circumstances}, widen
the period distribution sufficiently. The wide range of orbital
periods of very young and old binary systems is therefore a result of
cloud fragmentation and immediate subsequent magneto-hydrodynamical
processes operating within the multiple proto-stellar system.
\end{abstract}

\vskip 5mm
%
\keywords{binaries: general -- stars: formation -- stars: late-type --
open clusters and associations: general -- methods: n-body
simulations}

\section{INTRODUCTION} 
\label{sec:intro}
\noindent
The distribution of orbital parameters of binary systems pose
important constraints on the theory of star-formation. In particular,
the distribution of eccentricities and periods of late-type
Galactic-field binary systems are sufficiently well observed to
address the issue of their origin. 

The observed eccentricity distribution is approximately thermal
($f_{\rm e} \sim 2\,e$, with $f_{\rm e}\,de$ being the number of
orbits in the interval $e$ to $e+de$) for binaries with periods
$P\simgreat10^3$~days, but for systems with $P\simless 10^3$~days, $e$
and $P$ are correlated such that smaller $P$ imply, on average,
smaller $e$.  For Galactic late-type dwarfs the distribution of
periods, $f_{\rm P}$, is log-normal in $P$, with the notable feature
that $P$ ranges from 1 to $10^{10}$ days with a mean
log$_{10}P\approx4.8$ (fig.~1 in Kroupa 1995a).  Pre-main sequence
binaries show the same wide range of parameters and correlations
(Mathieu 1994).

The available cloud-collapse calculations have not been able to
produce the wide range of observed periods, and in particular do not
lead to short-period ($P\simless 10^3$~d) systems (Bodenheimer et
al. 2000 for a review). However, even the most recent numerical
refinements cannot describe stellar formation to the point where
gas-dynamical processes can be neglected, so that the final theoretical
orbital parameters of binary systems cannot be quantified. We do not
know for sure yet if there exists some significant
magneto-hydrodynamical mechanism that is important in transforming a
theoretical period distribution that results from fragmentation to the
final wide $f_{\rm P}$ observed already among~1~Myr old
populations. Alternatively, it may be possible that no such mechanism
is needed, and that the final (observed) orbital parameters of
Galactic-field systems are a result of gravitational encounters in
very dense embedded clusters that disperse rapidly.

It has already been demonstrated that stellar-dynamical interactions
in typical embedded clusters spanning a wide range of densities cannot
significantly widen a period distribution. Too few orbits are
redistributed to $P<10^3$~days by encounters assuming the primordial
period distribution is confined to the range $10^{3}-10^{7.5}$~days.
Similarly, typical embedded clusters cannot evolve an arbitrary
eccentricity distribution to the thermal form (Kroupa 1995b).

The purpose of the present paper is to study the evolution of a range
of {\it extremely dense} clusters to answer the question once and for
all whether encounters within a dense cluster environment can
significantly contribute to the observed width of $f_{\rm P}$.
Section~\ref{sec:code} briefly describes the codes used for the
$N$-body computations and the data reduction, and the initial
conditions. The results are presented in Section~\ref{sec:res}, and
the conclusions follow in Section~\ref{sec:conc}.

\section{THE CODES AND INITIAL CONDITIONS}
\label{sec:code}

\subsection{The $N$-body programme}
\noindent
A direct $N$-body code must deal efficiently with a range of dynamical
time-scales spanning many orders of magnitude, from days to hundreds
of~Myrs.  The code of choice is Aarseth's {\sc Nbody6} (Aarseth
1999). Special mathematical techniques are employed to transform the
space-time coordinates of closely interacting stars, which may be
perturbed by neighbours, such that the resulting equations of motion
of the sub-system are regular (Mikkola \& Aarseth 1993).
State-of-the-art stellar evolution is incorporated (Hurley, Pols \&
Tout 2000), as well as a standard Galactic tidal field (Terlevich
1987).

The velocity and position vectors of any individual centre-of-mass
particle (e.g. a star or binary system) diverges exponentially from
the true trajectory, through the growth of errors in $N$--body
computations (e.g. Goodman, Heggie \& Hut 1993). However, statistical
results from $N$--body calculations correctly describe the overall
dynamical evolution, as shown, for example, by Giersz \& Heggie (1994)
who compare ensembles of $N$-body computations with statistical
stellar-dynamical methods. Thus, for each model constructed here an
ensemble is created in order to obtain reliable estimates of the
relevant statistical quantities.

The output from {\sc Nbody6} is analysed by a software package that
finds all bound binary systems, and allows the construction of
distribution functions of orbital parameters, among many other things.

\subsection{Initial binary systems}
\noindent
Initial stellar masses are distributed according to a three-part
power-law IMF (Kroupa 2001b), $\xi(m)\propto m^{-\alpha}$, where
$\alpha=0.3$ for $0.01\le m <0.08\,M_\odot$, $\alpha=1.3$ for $0.08\le
m<0.5\,M_\odot$ and $\alpha=2.3$ for $0.5\le m/M_\odot$, and
$\xi(m)\,dm$ is the number of stars in the mass range $m$ to $m+dm$.

The total binary proportion is 
\begin{equation}
f_{\rm tot} = {N_{\rm bin}\over N_{\rm bin}+N_{\rm sing}},
\label{eq:f}
\end{equation}
where $N_{\rm bin}$ and $N_{\rm sing}$ are the number of binary and
single-star systems, respectively.  Initially, all stars are assumed
to be in binary systems ($f_{\rm tot}=1$) with component masses $m_1,
m_2$ chosen randomly from the IMF.  

The initial logarithmic binary-star periods are distributed uniformly
in the narrow interval $4.5 \le {\rm log}_{10}P\le 5.5$, $P$ in days.
The period distribution of late-type systems is
\begin{equation}
f_{\rm P} = {N_{\rm bin,P,lt}\over N_{\rm bin,lt}+N_{\rm sing,lt}},
\label{eq:fp}
\end{equation}
where $N_{\rm bin,P,lt}$ is the number of binaries with orbits in the
interval log$_{10}P$ and log$_{10}P+d{\rm log}_{10}P$ in which the
primary has a mass $0.08\le m\le1.5\,M_\odot$, and $N_{\rm sing,lt},
N_{\rm bin,lt}$ are the number of single stars and binaries with
primaries, respectively, in this mass interval.  The initial ($t=0$)
period distribution is given by
\begin{eqnarray}
f_{\rm P} = \left\{ \begin{array}{r@{\quad:\quad}l}
             0.5 & 4.5\le{\rm log}_{10}P\le5.5, \\
	     0.0 & {\rm otherwise},
	             \end{array} \right. 
\label{eq:finit}
\end{eqnarray}
the period distribution being constructed using decade intervals in
$P$ (see Fig.~\ref{fig:fp} below).  All binary systems initially have
orbits with eccentricity $e=0.75$, but the results are not sensitive
to this value (Kroupa 1995b).

These assumptions allow us to test the hypothesis that binary--binary
and binary--single-star encounters in very compact young clusters
widen the period distribution to the form observed in the Galactic
field, and re-distribute the eccentricities to give the thermal
distribution observed for Galactic-field systems.

\subsection{Cluster models}
\noindent
Clusters initially in virial equilibrium (VE) and in cold collapse
(CC) with $N=100$ and~1000 stars are set up to cover a range of
extreme initial conditions (Table~\ref{tab:mods}).

The VE models are spherical Plummer number-density profiles (Aarseth,
H\'enon \& Wielen 1974) initially, with position and velocity vectors
not correlated with the system masses. The initial half-mass radius,
$R_{0.5}$, is chosen to give a three-dimensional velocity dispersion,
$\sigma_{\rm 3D}$, that equals the orbital velocity of a binary with a
system mass of $1\,M_\odot$ and a period, $P=10^5$~days, near the
maximum in the G-dwarf period distribution of Duquennoy \& Mayor
(1991).  The central densities of these models are extreme and
probably unrealistic, in that many of the binary systems overlap.
However, these models allow us to study the very extreme situation in
which binary--binary interactions dominate initial cluster evolution,
and will thus pose limits on the possible widening of the period
distribution as a result of these interactions.

The CC models are uniform spheres, each cluster having no initial
velocity dispersion, with the position vectors and system masses being
uncorrelated. The increased binary--binary interactions near maximum
collapse may widen the period distribution to the observed values. 

\begin{table}
{\small
\begin{minipage}[t]{9cm}
\vskip 5mm

\begin{center}
\begin{tabular}{cccccccccc}

model  &$N$    &$R_{0.5}$   &$<\!\!m\!\!>$ 
&$\sigma_{\rm 3D}$ &log$_{10}t_{\rm cr}$ &log$_{10}\rho_{\rm C}$
&$M_{\rm cl}$   &$N_{\rm run}$ &Dynamical state\\
\tableline

       &       &[pc]        &[$M_\odot$]           
&[km/s]            &[Myr]                &[stars/pc$^3$] 
&[$M_\odot$]    &              &\\
\tableline

N2v    &$10^2$ &0.00287     &0.37
&4.59              &$-2.90$             &9.4
&37             &10            &VE\\

N2v1   &$10^2$ &0.0040      &0.30
&3.87              &$-2.68$             &8.9
&30             &10            &VE\\

N3v    &$10^3$ &0.0287      &0.37
&4.59              &$-1.90$             &7.4
&370            &5             &VE\\

N2c    &$10^2$ &1.2        &0.37
&0.0               &$+0.69$             &1.1
&37             &10            &CC\\

N3c    &$10^3$ &1.2        &0.37
&0.0               &$+0.37$             &2.1
&37             &5             &CC\\

\tableline

\end{tabular}
\end{center}
\end{minipage}
}
\caption{\small{ Initial cluster models. $N$ is the total number of
stars, and $M_{\rm cl}$ the cluster mass.  $R_{0.5}$ is the half-mass
radius of the Plummer-density distribution for the models that are
initially in virial equilibrium (VE). It is the radius of the
homogenous sphere for the models that undergo cold collapse (CC) that
have a vanishing initial three-dimensional velocity dispersion
$\sigma_{\rm 3D}$. The central density is $\rho_{\rm C}$.  All models,
except N2v1, have stars in the mass range $0.01-50\,M_\odot$ giving a
mean stellar mass $<\!\!m\!\!>=0.37\,M_\odot$. Model N2v1 contains
only stars with $0.08\le m \le1.1\,M_\odot$ having
$<\!\!m\!\!>=0.30\,M_\odot$. It contains no brown dwarfs and massive
stars, and thus provides comparison data for the more realistic other
cases.  No significant differences emerge.  In the VE models, $t_{\rm
cr}$ is the crossing time, whereas it is the time until maximum
contraction in the CC models. Each model is computed $N_{\rm run}$
times, each with a different random number seed.}}
\label{tab:mods}
\end{table}

\section{RESULTS} 
\label{sec:res}
\noindent

\subsection{Cluster evolution}
\begin{figure}
\begin{center}
\rotatebox{0}{\resizebox{0.77 \textwidth}{!}
{\includegraphics{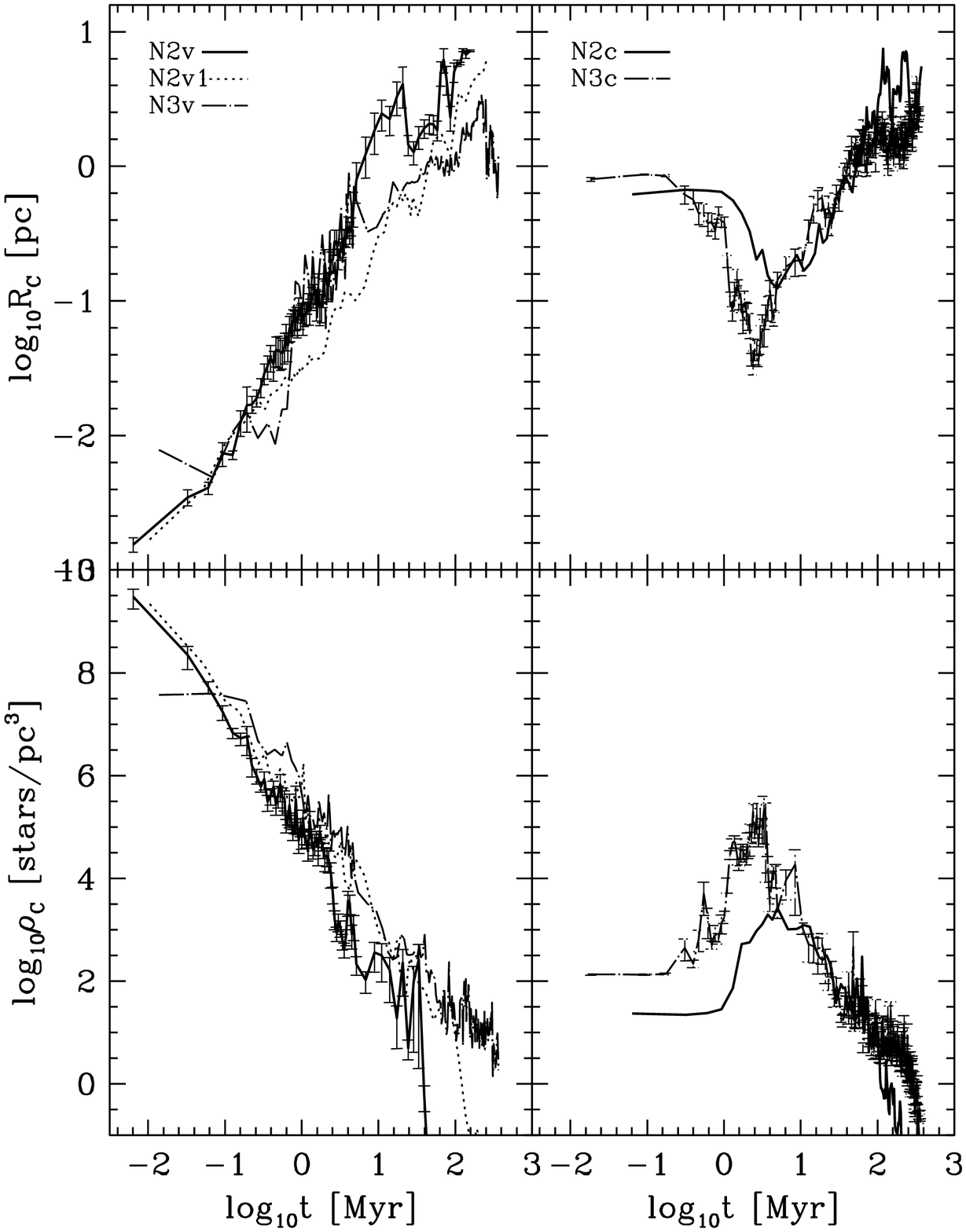}}}
\vskip 0mm
\caption
{\small{The core radius and central density. The scales are identical
in the top two and lower two panels. For clarity, the standard error of
the mean is shown for one model in each panel.  
}}
\label{fig:clev}
\end{center}
\end{figure}

\noindent
The evolution of the clusters is exemplified by evaluating the core
radius, $R_{\rm C}(t)$ (e.g. Kroupa, Aarseth \& Hurley 2001).  This
quantity measures the degree of concentration of a cluster. The
central number density is the density within $R_{\rm C}$, counting all
stars and brown dwarfs.  The evolution of both quantities is shown in
Fig.~\ref{fig:clev}.

The core radius of the VE models increases immediately as a
consequence of binary-star heating for the highly concentrated N2v and
N2v1 models, whereas it decreases in the less concentrated N3v model
during the first few $t_{\rm cr}$ as a result of mass segregation.
When energy re-arrangement through mass segregation has ended, the
most massive stars having reached the centre, $R_{\rm C}$ also expands
in this model.  It is interesting to note that the increase follows a
power-law, $R_{\rm C}\propto t^{r_R}$ with $r_R\approx0.95$, in all
cases. The central density decreases as a power-law, $\rho_{\rm
C}\propto t^{r_\rho}$ with $r_\rho\approx-2.2$.  The decay in
$\rho_{\rm C}$ sets-in immediately in models N2v and N2v1, but is
retarded in model N3v as a result of the settling of the heavy stars.

The CC models contract homologously until local density fluctuations
have increased sufficiently to form one dominating potential well,
thus focusing the further radial flow (Aarseth, Lin \& Papaloizou
1988).  At this point $\rho_{\rm C}$ begins to increase until it
reaches a maximum which defines the time of maximum contraction,
$R_{\rm C}$ being smallest then. After violent relaxation the
compactified clusters evolve as the VE models, albeit with a time-lag
through the collapse.

The clusters can be taken to be dissolved when $\rho_{\rm C}\simless
5$~stars/pc$^3$. The `final' distributions of orbital parameters
discussed below are evaluated approximately at this time taking
account of all stars, when no further stellar-dynamical, or
stimulated, evolution of the binary population occurs.

\subsection{The binaries}
\begin{figure}
\begin{center}
\rotatebox{0}{\resizebox{0.77 \textwidth}{!}
{\includegraphics{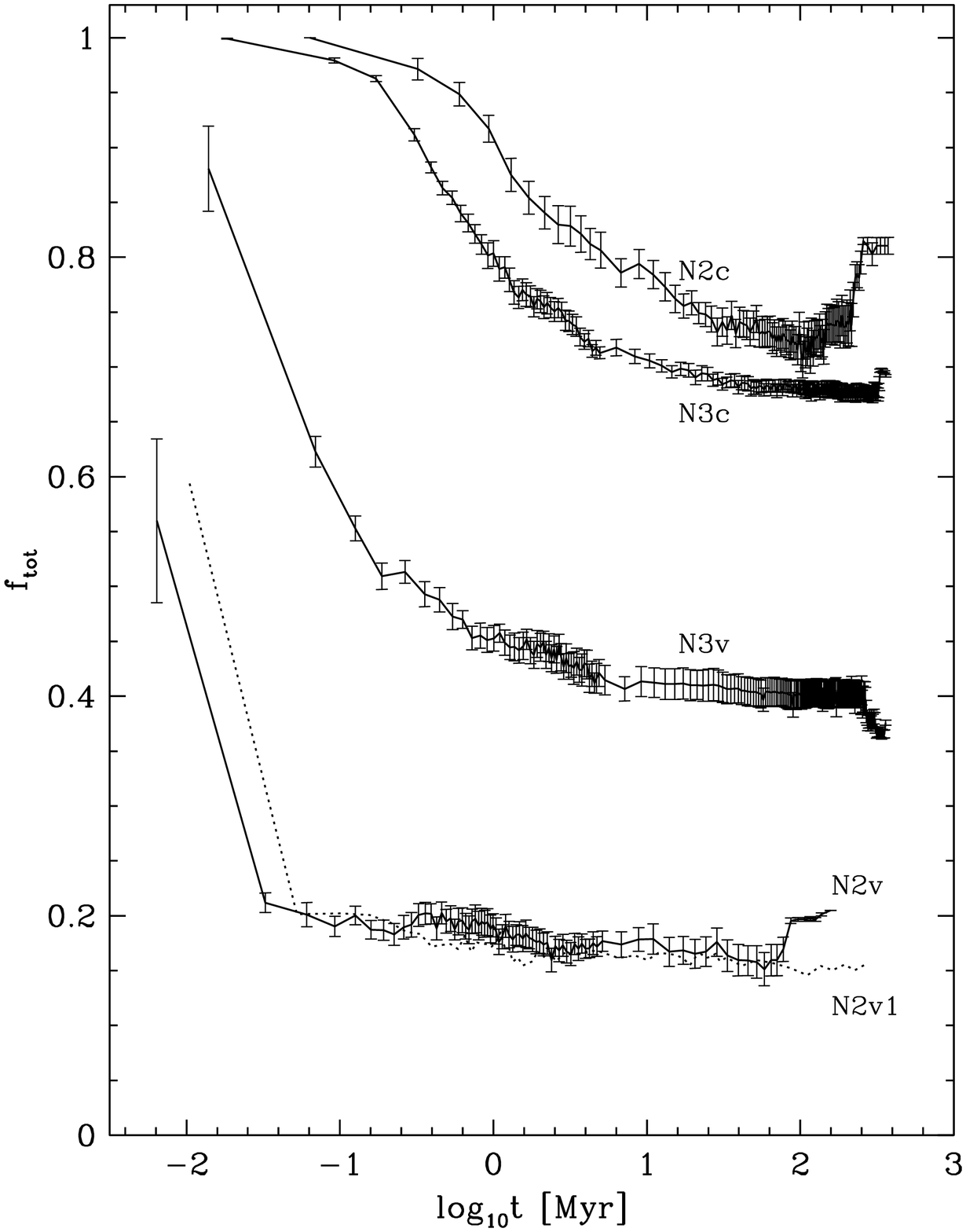}}}
\vskip 0mm
\caption
{\small{The evolution of the binary proportion
(eqn~\ref{eq:f}). Error-bars are standard error of the mean.
}}
\label{fig:ft}
\end{center}
\end{figure}

\noindent
The evolution of $f_{\rm tot}$ is plotted for all models 
in Fig.~\ref{fig:ft}. 

As noted above, the binaries initially overlap in the VE models,
causing immediate disruption of the widest systems leading to a
population of single stars in the cluster. Further binary depletion
occurs on a crossing time-scale. Meanwhile the cluster expands
(Fig.~\ref{fig:clev}), slowing further binary disruption, until it is
virtually halted at the point when the remaining binary population is
hard in the expanded cluster (Kroupa 2001a).  Fig.~\ref{fig:ft} shows
that finally $f_{\rm tot}\approx0.2$ for models N2v/v1, and $f_{\rm
tot}\approx0.4$ for model N3v. These values are too low in comparison
with the Galactic field ($f_{\rm tot}^{\rm obs}\approx0.6$), let alone
with pre-main sequence binaries ($f_{\rm tot}^{\rm obs}\approx0.8-1$).

In the CC models, binary disruption starts as soon as the cluster has
contracted sufficiently for the widest pairs to interact.  Disruption
continues throughout and beyond cold collapse, with binaries with an
increasing binding energy playing an increasing role.  During this
time, an increasing amount of kinetic energy is used up to disrupt
these binaries.  How the cold collapse is affected by such a large
primordial binary population is an interesting topic, but will not be
addressed further here. In both CC models, the final $f_{\rm tot}$ is
slightly higher than the binary proportion in the Galactic field.

\begin{figure}
\begin{center}
\rotatebox{0}{\resizebox{0.77 \textwidth}{!}
{\includegraphics{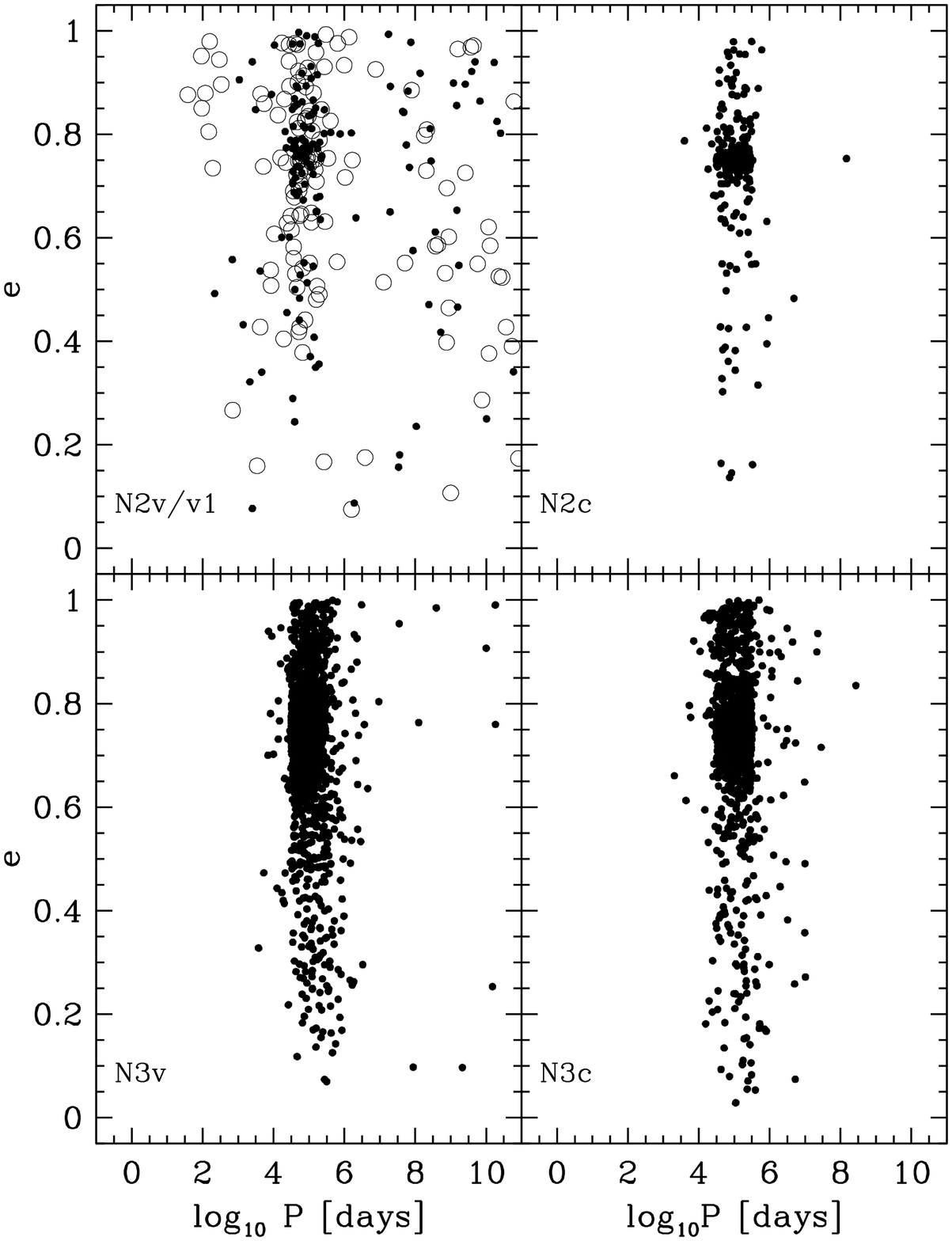}}}
\vskip 0mm
\caption
{\small{The final distributions in the eccentricity--period diagram.
The solid and open circles in the upper left panel are for model N2v
and N2v1, respectively.  The initial ($t=0$) distribution is given by
$e=0.75$ and $10^{4.5}\le P\le 10^{5.5}$.  }}
\label{fig:ep}
\end{center}
\end{figure}

During the binary--binary interactions some systems are perturbed or
exchange companions, rather than being disrupted. This leads to
changes in their orbital parameters. The final distributions of orbits
in the eccentricity--period diagram are shown in Fig.~\ref{fig:ep}.

A significant broadening of the eccentricity distribution is evident,
but the final distributions retain a maximum near $e=0.75$ which is
not consistent with the observational distribution (Duquennoy \& Mayor
1991). Thus, the stellar-dynamical encounters in extremely
concentrated clusters cannot evolve a delta-eccentricity distribution
to the observed thermal distribution. 

Similarly, a significant but not sufficient broadening of the period
distribution, $f_{\rm P}$, occurs in models N2v/v1, but the
effect on $f_{\rm P}$ is small in the other models.
\begin{figure}
\begin{center}
\rotatebox{-90}{\resizebox{0.77 \textwidth}{!}
{\includegraphics{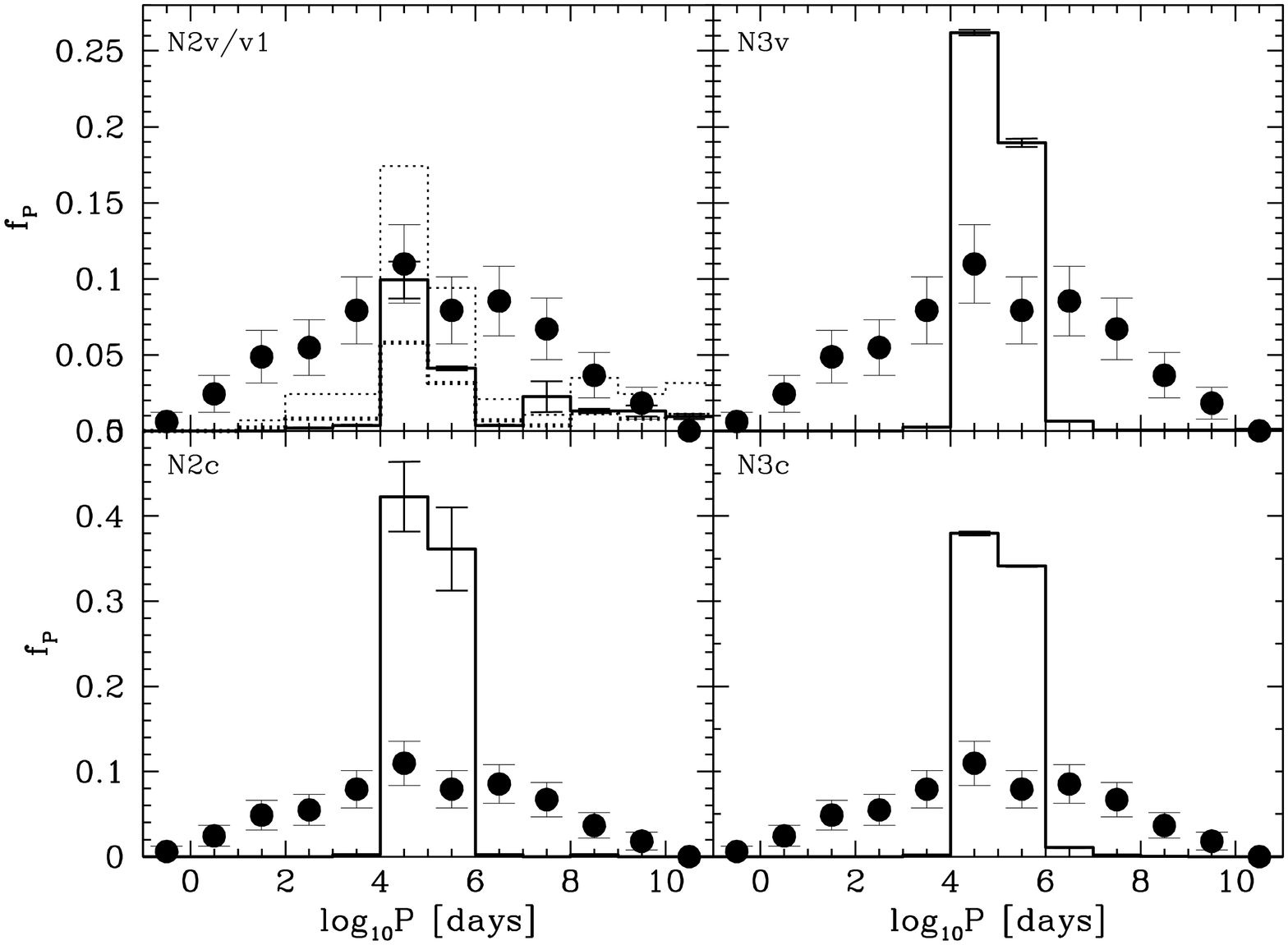}}}
\vskip 0mm
\caption
{\small{The final period distributions. For model N2v1, $f_{\rm P}$
and $3\times f_{\rm P}$ are plotted as thick and thin dotted lines,
respectively (see text).  The initial distributions are given by
eqn.~\ref{eq:finit}.  Filled circles show the G-dwarf period
distribution from Duquennoy \& Mayor (1992). M and K~dwarfs have
indistinguishable distributions (e.g. fig.~1 in Kroupa 1995a).  }}
\label{fig:fp}
\end{center}
\end{figure}
Fig.~\ref{fig:fp} demonstrates that the initial distribution cannot
evolve through stellar-dynamical interactions to the observed broad
log-normal distribution. This is the case even under the extreme
assumption that the evolution of the N2v1 cluster is halted
pre-maturely, for example through the expulsion of the remnant natal
gas, and that the single-star population that emerges from the
disruption of the binaries is lost preferentially through ejection and
mass segregation before gas-expulsion. In this thought experiment,
$f_{\rm P}$ will appear enhanced, since $f_{\rm P}=\eta\,N_{\rm
bin,P,lt}$ with $\eta=(N_{\rm bin,lt}+N_{\rm sing,lt})^{-1}$
(eqn~\ref{eq:fp}). In Fig.~\ref{fig:fp}, $f_{\rm P}'=3\times f_{\rm
P}$ is plotted for model N2v1, but even this artificially enhanced
period distribution disagrees with the observed distribution.

\section{CONCLUSION}
\label{sec:conc}
\noindent
The finding is thus that stellar-dynamical interactions in compact
star clusters cannot change an initially delta-eccentricity
distribution and narrow period distribution to the thermal
eccentricity and wide log-period distribution observed in the Galactic
field. It follows that the main characteristics of these distributions
(thermal $f_{\rm e}$ and orbits ranging from days to Myrs) must be a
result of fragmentation during star-formation and subsequent
magneto-hydrodynamical processes.

\newpage
\acknowledgements 
\vskip 10mm
\noindent{\bf Acknowledgements}
\vskip 3mm
\noindent 
The $N$-body calculations were performed on computers at the Institute
for Theoretical Astrophysics, Heidelberg University, using a variant
of Aarseth's {\sc Nbody6} code.

%

\end{document}